\documentclass[preprint,12pt]{elsarticle}

\usepackage[multidot]{grffile} 

\usepackage{amssymb}
\usepackage[separate-uncertainty=true, multi-part-units=single]{siunitx}

\usepackage{hyperref} 

\journal{Journal of Crystal Growth}

\begin{document}

\begin{frontmatter}



\title{Measurement of the Capillary Length for the Dendritic
       Growth of Ammonium Chloride}


\author{A.~J.~Dougherty\corref{cor1}}
\ead{doughera@lafayette.edu}
\ead[url]{http://sites.lafayette.edu/doughera}
\cortext[cor1]{Corresponding author}
\address{Department of Physics,
        Lafayette College, Easton, Pennsylvania, USA 18042-1789}

\begin{abstract}
We report the results of a new method of measuring the capillary
length for the dendritic crystal growth of non-faceted materials.
This method uses a nearly spherical crystal held near unstable
equilibrium in an oscillating temperature field.  For the growth of
ammonium chloride crystals from aqueous solution, previous published
estimates of the capillary length varied by over a factor of 20.
With this new method,
we find the product of the diffusion constant and capillary length $D d_0$
to be $\SI{0.78 \pm 0.07}{\micro m^3/s}$, similar
to that obtained for ammonium bromide crystals.

\end{abstract}


\end{frontmatter}


\section{Introduction}
Dendritic crystal growth is a common form of solidification observed
for the diffusion-limited growth of non-faceted materials.
It is important technologically because many metals and metal alloys
solidify in dendritic patterns, but it is also of more general
interest as one of the standard examples of pattern formation in a
nonlinear, nonequilibrium system \cite{Langer1980}.  For
reviews, see
Boettinger \textit{et al.}\ \cite{Trivedi2000},
Glicksman and March \cite{Glicksman-rev}, and
Huang and Wang \cite{Huang2012},

In the standard model of diffusion limited crystal growth, there are two
characteristic length scales \cite{Langer1980}.  One is the diffusion
length $l$, given by $l = 2 D / v$, where $D$ is the relevant diffusion
constant and $v$ is interface velocity.
The other is the capillary length, $d_0$, which is related
to the energy of the solid-liquid interface.
For many theories of dendritic crystal growth
\cite{Langer1980,Trivedi2000,Glicksman-rev,Kessler1988,BenAmar1993a}
the crystal tip growth speed $v$ and radius of curvature $\rho$
are related to those length scales by the quantity
$\sigma^\star$ defined by
\[
\sigma^\star = \frac{2 D d_0}{v \rho^2} .
\]
Experimental tests of those theories require measurements
of the relevant material parameters.

For a pure system, $d_0$ is given by
\begin{equation}
d_0 = \gamma T_M c_p / L^2
\label{eqn:d0:therm}
\end{equation}
where $\gamma$ is the solid-liquid surface tension, $T_M$ is the melting
temperature for a flat interface, $c_p$ is the specific heat per unit
volume, and $L$ is the latent heat per unit volume.

For growth from solution, $d_0$ is given by
\begin{equation}
d_0 = \frac{\gamma}{(\Delta C)^2 (\partial \mu / \partial C)}
\label{eqn:d0:chem}
\end{equation}
where $\Delta C$ is the difference in concentration between the solid
and liquid phases, and $\mu$ is the chemical potential.

For a few simple pure materials, such as helium \cite{Rolley1994}
or xenon \cite{Bilgram1993}, it is possible to measure the materials
constants in Eq.~\ref{eqn:d0:therm} directly and thus determine $d_0$.
More commonly, however, $d_0$ must be determined from measurements of
the solid-liquid interface shape under well-controlled experimental
conditions.

For pure materials, or for dilute solutions, Schaefer,
Glicksman, and Ayers \cite{Schaefer1975} showed how to use measurements
of the
curvature of the solid-liquid interface near a grain boundary to determine
$d_0$, and applied the technique to succinonitrile.
Singh and Glicksman \cite{Singh1989}
applied the same technique to pivalic acid.  Similar measurements have
been made for xenon \cite{Stalder2003} and pyrene \cite{akbulut:123505}.
Luo, Strachan, and Swift~\cite{Luo2005} developed a method to use the
maximum supercooling obtainable for a pure system to determine the
solid-liquid interfacial energy, and applied it to the ice-water
system.

For growth from solution, Dougherty and Gollub \cite{Dougherty1988}
modeled the shrinking of a nearly spherical crystal of ammonium
bromide.  Tanaka and Sano \cite{Tanaka1992} and
Sawada \textit{et al.}\ \cite{Sawada1995} applied the
same technique to ammonium chloride.
Gomes, Falc\~ao, and Mesquita \cite{Gomes2001} used the instability of
a cellular interface to measure the capillary length for a
nematic-isotropic interface in liquid crystals.

For ammonium bromide, Dougherty and Gollub \cite{Dougherty1988}
estimated $d_0 = \SI{0.28 \pm 0.04}{nm}$.  For ammonium chloride, Tanaka and
Sano~\cite{Tanaka1992}
estimated a much larger value of $d_0 = \SI{1.59 \pm 0.06}{nm}$, while
Sawada \textit{et al.}\  \cite{Sawada1995} estimated a much smaller value of
$d_0 = \SI{0.065}{nm}$.

In this paper, we extend the method of Ref.~\cite{Dougherty1988} to
maintain a nearly spherical crystal of ammonium chloride close to a
state of unstable equilibrium, and extract a value for the product
$D d_0$ from the
quasistatic growing and shrinking of that crystal.  Our final result
of $D d_0 = \SI{0.78 \pm 0.07}{\micro m^3/s}$, is intermediate between
the two previous
results, and is also comparable to our previously-reported value for
ammonium bromide.

\section{Theory}

The theory for the diffusion-limited growth of an isotropic spherical crystal
in an isothermal solution is developed in Ref.~\cite{Langer1980}.
Briefly, the dimensionless concentration field $u$ is presumed to follow
the diffusion equation in the quasistatic limit,
\[
\nabla^2 u(r) = 0 .
\]
For a spherical crystal of radius $R$,
the boundary conditions are that
$u(R) = 2 d_0/ R$,
where $d_0$ is the capillary length, and
$u(\infty) = \Delta$,
where $\Delta$ is the dimensionless
supersaturation, given in Ref.~\cite{Langer1980} by
\[
\Delta(T) = \frac{\mu(T) - \mu_{eq}}{\Delta C (\partial \mu / \partial C)}.
\]
The normal velocity $v_n$ of the interface is then given by
\[
v_n = D \nabla u \cdot \hat{n} \,,
\]
where $D$ is the diffusion coefficient for NH$_4$Cl in aqueous solution.
The final result is that the radial growth velocity, $dR/dt$,
is given by
\begin{equation}
\frac{dR}{dt} = \frac{D}{R} \left( \Delta(T) - \frac{2 d_0}{R} \right).
\label{eqn:vR1}
\end{equation}
This equation can also be used to define the critical radius,
$R_c(T) = 2 d_0 /\Delta(T)$.  Crystals larger than $R_c$
will grow, while crystals smaller than $R_c$ will shrink.

Equation \ref{eqn:vR1} is valid for an isolated crystal, but
in the present experiments, the crystal is denser than the surrounding
fluid, and rests against the bottom of the cell.  The crystal does not
wet the glass---experiments where the crystal rolls show that the
crystal remains roughly spherical.  (One such event is visible in
Fig.~\ref{fig:Rvst} around $t = \SI{0.73}{hrs}$.)

The presence of the boundary does affect the diffusive growth, however.
Following Ref.~\cite{Dougherty1988}, we model the presence of the
wall by considering
a sphere of radius $R$ resting on the bottom of the cell,
which is taken as the $z = 0$ plane.
We then solve the quasistatic diffusion equation for the
dimensionless concentration field $u$,
subject to the boundary conditions that
$u = \Delta$ on the far boundaries of the system,
$u = 2 d_0 / R$ on the surface of the crystal, and
the vertical gradient of $u$ is zero on the bottom of the cell.
We estimate the growth rate by averaging
the radial velocity  $D \nabla u \cdot \hat{r}$
over the surface of the crystal.  The result is $(74.7\pm0.3)$\% of the
value obtained without the cell wall at $z = 0$.  We model this as an
effective diffusion constant $D_\mathrm{eff} = \epsilon D$, where
$\epsilon = 0.747$.

For estimating $\Delta(T)$, since the typical temperature
variations are less than $\SI{\pm 0.02}{\celsius}$ in this experiment,
we adopt a simple
linear model for the product $D_\mathrm{eff} \Delta(T)$, namely
\begin{equation}
D_\mathrm{eff} \Delta(T) = B (T_{eq} - T)
\label{eqn:Delta}
\end{equation}
where $T_{eq}$ is the temperature at which a flat interface would be in
equilibrium, and $B$ is a proportionality constant
to be determined empirically.

The final result is that the radial growth velocity, $dR/dt$,
is given by
\begin{equation}
\frac{dR}{dt} = \frac{B}{R} \left(T_{eq} - T\right) -
                \frac{2 \epsilon D d_0}{R^2} .
\label{eqn:vR}
\end{equation}
The diffusion coefficient is not independently measured in these
experiments, so we instead determine the product $D d_0$.

\section{Materials and Methods}

The experiments were performed with solutions of ammonium chloride
(Fisher Scientific, 99.99\%) in water (Fisher Scientific, HPLC grade,
filtered through a $\SI{0.1}{\micro \meter}$ filter).
The concentration was approximately 36\% NH$_4$Cl by weight,
for a saturation temperature of approximately \SI{66.7}{\celsius}.

\subsection{Temperature control}
The solution
was placed in a $40 \times 10 \times 2$~mm$^3$ glass spectrophotometer cell
sealed with a Teflon stopper held in place by epoxy.
A thermistor was mounted on the cell for direct readings of the
temperature near the crystal.
The cell was mounted in a massive copper block, surrounded by an insulated
outer aluminum block,
and placed on a microscope stage.  The entire apparatus was
enclosed in a temperature-controlled insulated plexiglas box.

Temperatures were all measured with US Sensor PT503J2 thermistors,
which have an accuracy of $\pm\SI{0.2}{\celsius}$ and a sensitivity
of $\SI{-0.003}{\celsius/\ohm}$ in the temperature range of interest.
A thermistor was mounted inside the copper block and connected to a
Keithley 2000 6$\frac{1}{2}$ digit multimeter.  For heating, a Minco film
heater was mounted directly on the block and connected to an HP 6033A
power supply.  A similar system was connected to the outer aluminum
block.  The multimeters and power supplies were all connected to the
computer over the GPIB bus.  The temperature was controlled by a software
proportional-integral-derivative controller taking data every 2 seconds,
allowing complete programmatic control over the temperature during the
course of a run.  The rms temperature fluctuation of the copper block
around the target temperature was approximately $\SI{2e-4}{\celsius}$,
while the typical fluctuations in the outer aluminum block were less
than $\pm\SI{0.001}{\celsius}$.  The enclosing box was controlled by a
separate stand-alone temperature controller to $\pm\SI{0.05}{\celsius}$.

\subsection{Imaging}

\begin{figure}[htb!]
\includegraphics[width=0.45\textwidth]{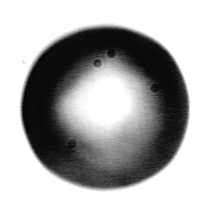}
    \caption{Crystal of ammonium chloride held in unstable
    equilibrium for 4 days.  The radius of the crystal is 26 $\mu$m.}
    \label{fig:img}
\end{figure}

Images were obtained from a charged coupled device (CCD) camera attached
to the microscope and acquired directly into the computer via a Data
Translation DT3155 frame grabber with a resolution of $640 \times 480$
pixels.  The ultimate resolution of the images was
$\SI{0.314 \pm 0.005}{\micro \meter/pixel}$.
A typical crystal image is shown in Fig.~\ref{fig:img}.

The interface position and radius of the nearly spherical crystals
were measured as in Ref.~\cite{Dougherty1991}.  To find the interface
position, the image intensity was first scanned on horizontal and vertical
lines passing through the crystal.  The intensity typically
dropped rapidly over the span of 4 or 5 pixels heading from outside to
inside the crystal.  In that transition region, a straight line was
fit to the intensity function, and the border was interpolated as the
position where that fit intensity equaled the average of the
intensity just outside and just inside the crystal.  The center of
the crystal was then approximated as the average of all the border
positions.  Since the
intensity profile of the interface is more reproducible if the
scanline is normal to the interface, a second set of scans was done radially,
and the center re-computed.  Finally, the radial scan was repeated to
get the final estimates of the border positions, which were then
converted to polar coordinates $r(\theta)$ around the center.
For the nearly spherical crystals used in this work, the estimate
of the center position was further adjusted to minimize the
amplitude of the first mode in a Fourier Series approximation of
$r(\theta)$.

\subsection{Methods}

To obtain a single crystal, the solution was heated to dissolve all the
NH$_4$Cl, stirred to eliminate concentration gradients, and then cooled
to initiate growth.  Many crystals would nucleate, but upon warming,
most would dissolve.  The system was again stirred to bring the largest
remaining crystals into the field of view, and an automated process was
initiated to acquire images, measure their size, and slowly adjust
the temperature until only a single isolated crystal remained.

Specifically, at each time step (typically \SI{20}{s}) we measured the
radius $R$ and used it and the preceding 3 measurements to estimate the
growth velocity and hence the critical radius $R_c$.  (This required a
preliminary estimate of $D d_0$, but the result was not particularly
sensitive to the precise value.)  Although Eq.~\ref{eqn:vR1} is only
valid for a single isolated spherical crystal, we found that it provided
a workable approximation even for irregularly shaped crystals and even
for small collections of crystals, provided the overall velocity was
relatively small $( \ll \SI{1}{\micro m/s})$.  For larger velocities,
where the calculated $R_c$ values were not reasonable, we made
ad-hoc estimates of plausible values.  We then compared $R$ to $R_c$
and to the desired target size $R_\mathrm{target}$, and adjusted the
temperature up or down proportionally to the difference $R - (R_c +
R_\mathrm{target})/2$.

This process was continued until the largest crystal was the desired
size.  Smaller crystals would dissolve, since they all had $R < R_c$.
Depending on the initial conditions, the whole process typically took
about a week.  Although such a spherical crystal is in a state of unstable
equilibrium, we found it possible to maintain it indefinitely, provided we
continually monitored the size and adjusted the temperature accordingly.
This isolated, nearly spherical, crystal was allowed to stabilize for
several days.  An example of a crystal held stable for 4 days is shown
in Fig.~\ref{fig:img}.

Once a single, stable crystal was obtained, the temperature of the
sample was slowly oscillated with an amplitude of $\SI{0.015}{\celsius}$ and
a period of either \SI{6000}{s} or \SI{8000}{s}.
This caused the crystal to alternatively grow and
dissolve slowly, but the size and growth rate
changed enough that both terms on the right of Eq.~\ref{eqn:vR} were
significant.

We then used the data for $R(t)$ and $T(t)$ to numerically integrate Eq.~\ref{eqn:vR}
and compare to the measured $R(t)$.  We determined the constants
$B$, $T_{eq}$, and $D d_0$ in Eqs.~\ref{eqn:vR} and \ref{eqn:Delta}
by minimizing the squared difference between the integrated
prediction and the
measured values for $R(t)$.

\section{Results}

\begin{figure}[htb!]
\includegraphics[width=0.5\textwidth]{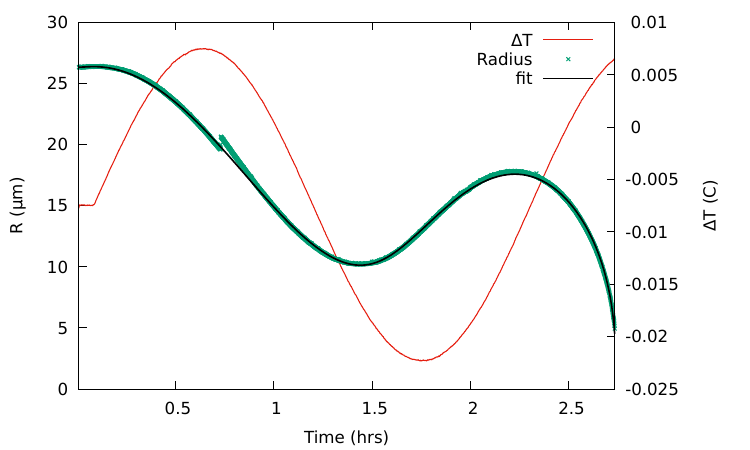}
    \caption{Measured (dots) and fit (line) values for the radius as a
    function of time of an oscillating crystal.  The driving temperature
    variations $\Delta T = T - T_{eq}$ are also shown (red online).
    The crystal rolled slightly around $t = \SI{0.73}{hours}$.
    }
    \label{fig:Rvst}
\end{figure}

The resulting fit is shown in Fig.~\ref{fig:Rvst}.
The best fit parameters are
$D d_0 = \SI{0.78 \pm 0.07}{\micro m^3/s}$,
$B  = \SI{6.36 \pm 0.10}{\micro m^2/s \celsius}$, and
$T_{eq} = \SI{66.727 \pm 0.001}{\celsius}$,
where the uncertainties are the standard deviation of the mean over
three independent runs.

Comparisons with previous measurements for $d_0$ require an estimate
of the diffusion coefficient $D$.
For lower concentrations and temperatures, the diffusion coefficient was
measured by Lutz and Mendenhall \cite{Lutz2000} and by Hall, Wishaw,
and Stokes \cite{Hall1953}.  Extrapolating those published values to the
temperatures and concentrations used in this work,
we estimate $D = \SI{2500 \pm 200}{\micro m^2/s}$, similar to the values of
$\SI{2600}{\micro m^2/s}$ used by Tanaka and Sano \cite{Tanaka1992},
and
$\SI{2280}{\micro m^2/s}$ used by Sawada \textit{et al.}\ \cite{Sawada1995}.
With this value of $D$, we estimate
$d_0 = \SI{0.31 \pm 0.04}{\nano m}$ for NH$_4$Cl.  This
value is in between the values of
$d_0 = \SI{1.59 \pm 0.06}{nm}$ and $d_0 = \SI{0.065}{nm}$
reported in Refs.~\cite{Tanaka1992} and \cite{Sawada1995}.
It also compares favorably with the value of
$\SI{0.28 \pm 0.04}{nm}$
previously reported for NH$_4$Br \cite{Dougherty1988},

\section{Conclusion}
There were two main factors we found to be important in getting consistent
results from this experiment.  First, control and measurement of the
temperature at the sample was critical.  In preliminary trials with poorer
insulation, or where the outer aluminum block temperature fluctuated more,
we were unable to get satisfactory fits
to Eq.~\ref{eqn:vR} over long time scales as in Fig.~\ref{fig:Rvst}.

Second, the oscillating protocol proved a robust probe of
Eq.~\ref{eqn:vR}.  In previous work on slowly dissolving crystals, as
in Ref.~\cite{Dougherty1988}, we typically found that the fitted value
for $d_0$ depended very sensitively on the last few data points, where
the crystal was shrinking rapidly just before completely dissolving.
Small drifts in temperature on the order of $\SI{\pm 0.001}{\celsius}$ could
make significant differences in the quality of the fit.  In Ref.\
\cite{Tanaka1992}, the fit does not include the last two data points.
By contrast, Fig.~\ref{fig:Rvst} includes
both growing and dissolving segments so that both terms in Eq.~\ref{eqn:vR}
were significant, and the fit was well-constrained.

One remaining uncertainty is the possible role of convection in
this experiment.  The density differences caused by the growth and
dissolution of the crystal could drive small convection currents.
However, the excellent fit to the data suggests that Eq.~\ref{eqn:vR}
still provides an effective model for the growth process.

\section*{Acknowledgments}
The author thanks F.\ Scott Stinner for his work on an earlier version
of this experiment, and Lafayette College for financial support.

This research did not receive any specific grant from funding agencies
in the public, commercial, or not-for-profit sectors.

Data used in this paper is available through Mendeley at
\url{https://data.mendeley.com/datasets/kr2mgbymr3/1}.

Declarations of interest: none.



\section*{References}
\bibliographystyle{elsarticle-num}
\bibliography{caplen}

\begin{thebibliography}{10}
\expandafter\ifx\csname url\endcsname\relax
  \def\url#1{\texttt{#1}}\fi
\expandafter\ifx\csname urlprefix\endcsname\relax\def\urlprefix{URL }\fi
\expandafter\ifx\csname href\endcsname\relax
  \def\href#1#2{#2} \def\path#1{#1}\fi

\bibitem{Langer1980}
J.~S. Langer, Instabilities and pattern formation in crystal growth, Reviews of
  Modern Physics 52~(1) (1980) 1--28.
\newblock \href {http://dx.doi.org/10.1103/RevModPhys.52.1}
  {\path{doi:10.1103/RevModPhys.52.1}}.

\bibitem{Trivedi2000}
W.~J. Boettinger, S.~R. Coriell, A.~L. Greer, A.~Karma, W.~Kurz, M.~Rappaz,
  R.~Trivedi, Solidification microstructures: {{Recent}} developments, future
  directions, Acta Mater. 48~(1) (2000) 43--70.
\newblock \href {http://dx.doi.org/10.1016/S1359-6454(99)00287-6}
  {\path{doi:10.1016/S1359-6454(99)00287-6}}.

\bibitem{Glicksman-rev}
M.~E. Glicksman, S.~P. March, in: D.~J.~T. Hurle (Ed.), Handbook of Crystal
  Growth, Elsevier Science, 1993, p. 1081.

\bibitem{Huang2012}
W.~Huang, L.~Wang, Solidification researches using transparent model materials
  \textemdash{} {{A}} review, Science China Technological Sciences 55~(2)
  (2012) 377--386.
\newblock \href {http://dx.doi.org/10.1007/s11431-011-4689-1}
  {\path{doi:10.1007/s11431-011-4689-1}}.

\bibitem{Kessler1988}
D.~A. Kessler, J.~Koplik, H.~Levine, Pattern {{Selection}} in {{Fingered Growth
  Phenomena}}, Adv. Phys. 37~(3) (1988) 255--339.
\newblock \href {http://dx.doi.org/10.1080/00018738800101379}
  {\path{doi:10.1080/00018738800101379}}.

\bibitem{BenAmar1993a}
M.~Benamar, E.~Brener, Theory of {{Pattern Selection}} in 3-{{Dimensional
  Nonaxisymmetric Dendritic Growth}}, Phys. Rev. Lett. 71~(4) (1993) 589--592.
\newblock \href {http://dx.doi.org/10.1103/PhysRevLett.71.589}
  {\path{doi:10.1103/PhysRevLett.71.589}}.

\bibitem{Rolley1994}
E.~Rolley, S.~Balibar, F.~Graner, Growth shape of $^{3}\mathrm{He}$ needle
  crystals, Phys. Rev. E 49 (1994) 1500--1506.
\newblock \href {http://dx.doi.org/10.1103/PhysRevE.49.1500}
  {\path{doi:10.1103/PhysRevE.49.1500}}.

\bibitem{Bilgram1993}
J.~H. Bilgram, E.~Hurlimann, Dendritic solidification of rare-gases, Progress
  in Crystal Growth and Characterization of Materials 26 (1993) 67--86.

\bibitem{Schaefer1975}
R.~Schaefer, M.~Glicksman, J.~Ayers, High-{{Confidence Measurement}} of
  {{Solid}}-{{Liquid Surface}}-{{Energy}} in a {{Pure Material}}, Philosophical
  Magazine 32~(4) (1975) 725--743.
\newblock \href {http://dx.doi.org/10.1080/14786437508221616}
  {\path{doi:10.1080/14786437508221616}}.

\bibitem{Singh1989}
N.~B. Singh, M.~E. Glicksman, Determination of the {{Mean Solid}}-{{Liquid
  Interface Energy}} of {{Pivalic Acid}}, J. Cryst. Growth 98~(4) (1989)
  573--580.
\newblock \href {http://dx.doi.org/10.1016/0022-0248(89)90293-5}
  {\path{doi:10.1016/0022-0248(89)90293-5}}.

\bibitem{Stalder2003}
I.~Stalder, J.~H. Bilgram, The measurement of the solid-liquid surface free
  energy of xenon, J. Chem. Phys. 118~(17) (2003) 7981--7984.
\newblock \href {http://dx.doi.org/10.1063/1.1565319}
  {\path{doi:10.1063/1.1565319}}.

\bibitem{akbulut:123505}
S.~Akbulut, Y.~Ocak, U.~Boyuk, M.~Erol, K.~Keslioglu, N.~Marasli, Solid-liquid
  interfacial energy of pyrene, J.\ Appl.\ Phys. 100~(12) (2006) 123505.
\newblock \href {http://dx.doi.org/10.1063/1.2402098}
  {\path{doi:10.1063/1.2402098}}.

\bibitem{Luo2005}
S.~N. Luo, A.~Strachan, D.~C. Swift, Deducing solid-liquid interfacial energy
  from superheating or supercooling: Application to {{H2O}} at high pressures,
  Model. Simul. Mater. Sci. Eng. 13~(3) (2005) 321--328.
\newblock \href {http://dx.doi.org/10.1088/0965-0393/13/3/002}
  {\path{doi:10.1088/0965-0393/13/3/002}}.

\bibitem{Dougherty1988}
A.~Dougherty, J.~P. Gollub, Steady-state dendritic growth of
  {{NH}}{\textsubscript{4}}{{Br}} from solution, Physical Review A 38~(6)
  (1988) 3043--3053.
\newblock \href {http://dx.doi.org/10.1103/PhysRevA.38.3043}
  {\path{doi:10.1103/PhysRevA.38.3043}}.

\bibitem{Tanaka1992}
A.~Tanaka, M.~Sano, Measurement of the kinetic effect on the concentration
  field of a growing dendrite, J.\ Cryst.\ Growth 125~(1–2) (1992) 59--64.
\newblock \href {http://dx.doi.org/10.1016/0022-0248(92)90320-I}
  {\path{doi:10.1016/0022-0248(92)90320-I}}.

\bibitem{Sawada1995}
T.~Sawada, K.~Takemura, K.~Shigematsu, S.~ichi Yoda, K.~Kawasaki, Diffusion
  field around a dendrite growing under microgravity, Phys. Rev. E 51 (1995)
  R3834--R3837.
\newblock \href {http://dx.doi.org/10.1103/PhysRevE.51.R3834}
  {\path{doi:10.1103/PhysRevE.51.R3834}}.

\bibitem{Gomes2001}
O.~A. Gomes, R.~C.~F. ao, O.~N. Mesquita, Anomalous capillary length in
  cellular nematic-isotropic interfaces, Phys. Rev. Lett. 86 (2001) 2577--2580.
\newblock \href {http://dx.doi.org/10.1103/PhysRevLett.86.2577}
  {\path{doi:10.1103/PhysRevLett.86.2577}}.

\bibitem{Dougherty1991}
A.~Dougherty, Surface-{{Tension Anisotropy}} and the {{Dendritic Growth}} of
  {{Pivalic Acid}}, J. Cryst. Growth 110~(3) (1991) 501--508.
\newblock \href {http://dx.doi.org/10.1016/0022-0248(91)90286-E}
  {\path{doi:10.1016/0022-0248(91)90286-E}}.

\bibitem{Lutz2000}
J.~L. Lutz, G.~D. Mendenhall, Diffusion coefficients by {{NMR}}-spin echo
  methods for the systems water-ammonium chloride, water-succinonitrile, and
  acetone-succinonitrile, J. Cryst. Growth 217~(1-2) (2000) 183--188.
\newblock \href {http://dx.doi.org/10.1016/S0022-0248(00)00504-2}
  {\path{doi:10.1016/S0022-0248(00)00504-2}}.

\bibitem{Hall1953}
J.~R. Hall, B.~F. Wishaw, R.~H. Stokes, The {{Diffusion Coefficients}} of
  {{Calcium Chloride}} and {{Ammonium Chloride}} in {{Concentrated Aqueous
  Solutions}} at 25$^\circ$, J. Am. Chem. Soc. 75~(7) (1953) 1556--1560.
\newblock \href {http://dx.doi.org/10.1021/ja01103a011}
  {\path{doi:10.1021/ja01103a011}}.

\end{thebibliography}

\end{document}